# Rough Set Model for Discovering Hybrid Association Rules

Anjana Pandey[1]
Research Scholar,
Maulana Azad National Institute of Technology Bhopal,(India)
apeeshukla@gmail.com

K.R.Pardasani[2]
Professor & Head
Department of Mathematics
Maulana Azad National Institute of Technology,Bhopal,(India)
krpardasani@hotmail.com

*Abstract*—In this paper, the mining of hybrid association rules with rough set approach is investigated as the algorithm RSHAR.The RSHAR algorithm is constituted of two steps mainly. At first, to join the participant tables into a general table to generate the rules which is expressing the relationship between two or more domains that belong to several different tables in a database. Then we apply the mapping code on selected dimension, which can be added directly into the information system as one certain attribute. To find the association rules, frequent itemsets are generated in second step where candidate itemsets are generated through equivalence classes and also transforming the mapping code in to real dimensions. The searching method for candidate itemset is similar to apriori algorithm. The analysis of the performance of algorithm has been carried out.

## I. INTRODUCTION

Association rule mining finds interesting association or correlation relationship among a large data set of items [1, 2]. The discovery of interesting association rules can help in decision making process. Association rule mining that implies a single predicate is referred as a single dimensional or *interdimension association rule* since it contains a single distinct predicate with multiple occurrences (the predicate occurs more than once within the rule). The terminology of *single dimensional or intradimension association rule* is used in multidimensional database by assuming each distinct predicate in the rule as a dimension. For instance, in *market basket analysis*,. In *market basket analysis*, it might be discovered a Boolean association rule "laptop $\Rightarrow$ b/w printer" which can also be written as a single dimensional association rule as follows [3]:

Rule-1
  buys(X, "laptop") $\Rightarrow$ buys(X, "b/w printer"),

where *buys* is a given predicate and *X* is a variable representing customers who purchased items (e.g. *laptop* and *b/w printer*). In general, *laptop* and *b/w printer* are two different data that are taken from a certain database attribute, called *items*. In general, *Apriori* [1] is used an influential algorithm for mining frequent itemsets for generating Boolean (single dimensional) association rules.

Additional relational information regarding the customers who purchased the items, such as customer age, occupation, credit rating, income and address, may also have a correlation to the purchased items. Considering each database attribute as a predicate, it can therefore be interesting to mine association rules containing *multiple* predicate, such as:

Rule-2:
Age ("20..29") ? sex("Male") ? income("5K..7K") ? buys("Laptop")

Where there are four predicates, namely age, sex, income and buys. Association rules that involve two or more dimensions or predicates can be referred to as *multidimensional association rules*. Multidimensional rules with no repeated predicates are called *interdimension association* rules (e.g Rule-2)[4] .On the other hand, multidimensional association rules with repeated predicates, which contain multiple occurrences of some predicates, are called *hybrid-dimension association rules* .The rules may be also considered as combination (hybridization) between intradimension association rules and interdimension association rules.This notion of hybrid association rules is a development of basic association rules, since it involves more complex rules and is more likely happen in real world data. An example of a hybrid association rule is the following, where the predicate *buys* is repeated

Rule-3
Times (1998) ? Location (Melb) ? Buy(Beer) ? Buy(Diaper)
{sup=30%,conf=80%}

This rule means that in the year 1998 customers in *Melbourne* who buy *beer* and buy *diaper* together support 30% of total transactions and those customers in *Melbourne* who buy *beer* have a confidence or probability of buying *diaper* together of 80%. are numbered with Roman numerals.
 This example uses three different types of predicate which are Times, Location and Buy where predicate *Buy* is repeated. Unlike normal association rules, it uses an only single predicate which is predicate *Buy*. The formal model of a hybrid association rule is similar to a normal association rule, although a hybrid association rule has to show its predicate's type as well.

Here we discuss hybrid association rules in transaction database. The structure of the paper is the following. Section 2 describes data preparation for the further process of generation rules. Here we will discuss a process of joining table from database. Afterthat relational schema has to be transformed into bitmap table. Section 3, presents the rough set model which is used in RSHAR. Section 4, introduces RSHAR algorithm for mining of hybrid association rules with rough set. Section 5 presents some performance result showing the effectiveness of our method. Finally, section 6 concludes the paper. 2. Background

## II. BACKGROUND

In this section we provide a short introduction of process of joining tables from relational database and concept of bitmap table which are used in our algorithm.

2.1 Method for joining of Tables
In general, the process of mining data for discovering association rules has to be started from a
single table (relation) as a source of data representing relation among item data. Formally, a relational data table [5] $R$ consists of a set of tuples, where $t_i$ represents the $i$-th tuple and if there are $n$ domain attributes $D$, then $t_i = (d_{i1}, d_{i2}, \cdots, d_{in})$. Here, $d_{ij}$ is an atomic value of tuple $t_i$ with the restriction to the domain $D_j$ where $d_{ij} \in D$. Formally, a relational data table $R$ is defined as a subset of the set of cross product $D_1 \times D_2 \times \cdots \times D_n$, where $D = \{D_1, D_2, \ldots, D_n\}$
Tuple $t$ (with respect to $R$) is an element of $R$. In general, $R$ can be shown in Table 1.

| Tuples | $D_1$ | $D_2$ | $\cdots$ | $D_n$ |
|---|---|---|---|---|
| $t_1$ | $d_{11}$ | $d_{12}$ | $\cdots$ | $d_{1n}$ |
| $t_2$ | $d_{21}$ | $d_{22}$ | $\cdots$ | $d_{2n}$ |
| $\vdots$ | $\vdots$ | $\vdots$ | $\ddots$ | $\vdots$ |
| $t_r$ | $d_{r1}$ | $d_{r2}$ | $\cdots$ | $d_{rn}$ |

Table1 A Relational Database

In many case the database may consist of several relational data tables in which they have relation one to each others. Their relation may be represented by Entities Relationship Diagram (ERD). Hence, suppose we need to process some domains (columns) data that are parts of different relational data tables, all of the involved tables have to be combined (joined) together providing a *general data table*. In the process of joining the tables, it is not necessary that all domains (fields) of the all combined tables have to be included in the targeting table. Instead, the targeting table only consists of interesting domains data that are needed in the process of mining rules. The process of joining tables can be performed based on two kinds of data relation as follows.
1. On the basis of Metadata
Information of relational tables can be stored in a metadata. Simply, a metadata can be represented by a table. Metadata can be constructed using the information of relational data by an Entity relationship Diagram (ERD). A detailed description of metadata and ERD can be found in inten[6]).
2. On the basis of function defined by the user
It is possible for user to define a mathematical function (or table) relation for connecting two or more domains from two different tables in order to perform a relationship between their entities. Generally, the data relationship function performs a mapping process from one or more domains from an entity to one or more domains from its partner entity. Four possibilities of function $f$ performing a mapping process are given by [6]
1) One to one relationship
$$f : C_i \to D_k$$
2) One to many relationship
$$f : C_i \to D_{p1} \times D_{p2} \times \ldots \times D_{pk}$$
3) Many to one relationship
$$f : C_{m1} \times C_{m2} \times \ldots \times C_{mk} \to D_k$$
4) Many to many relationship

$$f : C_{m1} \times C_{m2} \times \ldots \times C_{mk} \to$$
$$D_{p1} \times D_{p2} \times \ldots \times D_{pk}$$

2.2 Data Structure 'Bitmap'

In relation table some attributes has quantitative values which can be discretized as some categorical values on behalf of certain range. Then the form of information system is changed to that each attribute in the new database is an exact value of one item in original system, and each attribute value is either 1 or 0,expressing if it is present there is a '1', otherwise a '0' in the bitmap[7].
Example 1. For an attribute with no- binary domain. each attribute value corresponds to one item. for example, for attribute 'age' with domain(age)={young,middle,old} (i={1,2,3} the following items result:$A_1$ ="age_young",$A_2$="age_middle",$A_3$="age_old" (see fig.1)

| TID | Age |
|---|---|
| 1 | Young |
| 2 | Middle |
| 3 | Middle |
| .. | .. |

Transformation

| TID | $A_1$ | $A_2$ | $A_3$ |
|---|---|---|---|
| 1 | 1 | 0 | 0 |
| 2 | 0 | 1 | 0 |
| 3 | 0 | 1 | 0 |
| .. | . | .. | .. |

Fig 1.Transformation of relational data into an efficient bitmap representation for attributes with no-binary domains.

## III. ROUGH SET

In 1982 Z.Pawalak [8] introduced a new tool to deal with vagueness, called the "rough set". It is a method for uncovering dependencies in data, which are recorded by relations. The rough set philosophy is based on the idea of classification. A detailed introduction to rough set theory can be found in Munakata [9].

3.1 Model

The rough set method operates on data matrices, so called "Information System. It contains data about the universe $U$ of interest, condition attributes and decision attributes. The goal is to derive rules that give information how the decision attributes depend on the condition attributes. By an information system S, S= {U, *At, V, f*}, where U is a finite set of objects, U= { $x_1, x_2, \ldots, x_n$ }, *At* is a finite set of attributes, the attribute in *At* is further classified into two subsets, condition attributes C and decision attribute D. In Hybrid association rule condition attributes and decision attributes are not disjoint. Thus, a formal model of hybrid association rules is

$$d_1(val), d_2(val), \ldots, d_m(val) \rightarrow$$
$$d_2(val), \ldots, d_m(val)$$

V= $\bigcup_{p \in A} V_p$, and $V_p$ is a domain of attribute p.

Here the function $f$ performs a mapping code of $d_2(val), \ldots, d_m(val)$ into one simple attribute which can be added directly into the information system as one certain attribute, it will only posses one column in the information system, analogous
an item.
      A prerequisite for rule generation is a partitioning of $U$ in a finite number of blocks, so called equivalence classes[10], of same attribute values by applying an equivalence relation.

## IV. PROPOSED ALGORITHM

We propose two algorithms for mining of interdimension association rules in transaction database .Those algorithms are :CombineDims, and GenFI.

First we apply the CombineDims algorithm to combine the selected dimensions in order to provide the framework for mining hybrid association rules. Then, we apply the GenFI algorithm to discover frequent itemsets in the transaction database. For the new information system, the searching of frequent itemsets is easy based on the concept of equivalence class.

*4.1 CombineDims Algorithm*
We prepare the data from the general table as follows:

1. Select the dimension $d_2, \ldots, d_m$ From the general tables where ($d_2 = duser_2$) And…..($d_m = duser_m$).This syntax create an initialized table IntTab for mining multidimensional association rule. Now we apply one distinct mapping code which is stored on MapTab for selected dimension as follows.
(Times dimension, channels dimension/Products dimension, and mapping code)
    ('Jan 1998','Direct sales/Men-Jeans','0001')
Here we combine two dimensions: channels and products into one mapping code '0001'.

- Line 4 checks selected dimension: $\{d_2, d_3, \ldots d_m\}$ whether or not they already have its mapping code.
- Line 5 generates and stores a new mapping code for selected dimensions.
- Line 9,searches a mapping code in table MapTab for selected dimension in table IntTab

The following are the details of our proposed algorithms. Note that notations in table 2 are used for our proposed algorithms.
    Table 2. Notation

| Notation | Meaning |
|---|---|
| D | Sets of dimensions and its values $\{d_1, d_2, \ldots d_m\}$ |
| ComDim | Combine Dimensions and its values $\{d_2, \ldots d_m\}$ |
| IntTab | Initialize Table $\{D, count\}$ |
| MdTab | Md Table $\{d_1, MapCode\}$ |
| KeyTab | Key Table $\{d_1\}$ |
| MdTabProcess | Process Md; contains $\{d_1,$ List of MapCode$\}$ |
| TmpLargeTab | Temp Large Itemset Table {List of ComDim,Level,Sup} |

1.Procedure CombineDims
2. X={Total rows of table IntTab}
3.For I=1 to X Loop //on  table IntTab
4. If !CheckMapCode( $d_2, d_3 \ldots d_m$ ) then
5.    GenMapCode( $d_2, d_3 \ldots d_m$ );
6. End IF;
7. End Loop;
8.For J= 1 to X Loop// on table IntTab
9.S=FindMapCode( $d_2, d_3 \ldots d_m$ );
10. Insert MdTab(IntTab( $d_1.key$) ,MapTab(MapCode))
11. End Loop;

After creating MdTab, we use that table in the GenFI algorithm to discover frequent itemset on hybrid association rules in transaction database.Here are the details of the working of our algorithm.
- Line 6 checks whatever $d_1$ value exists on table KeyTab.
- Line 7 inserts a new record: $d_1$ value on table keyTab.
- Lines 11-15 create list of mapping codes taken from table MdTab and selected $d_1$ from KeyTab.

- Line 16 inserts a new record on table MdTabprocess.
- Line 18 creates all large itemset from table MdTabprocess with specified user minimum support and inserts the result into table TmpLargeTab.
- Line 19 changes the mapping code from table TmpLargeTab into the real dimension value. Thus all the large itemset after mapping the code are stored in table LargeTab

*4.2 GenFI Algorithm*

1. X={total rows of MdTab};
2. Y={total rows of table keyTab};//key table{ $d_1$ }
3. N={total attributes of selected $d_m.key$};
4. For I = 1 to X Loop // on table MdTab
5. IF !CheckKey($d_1$) then
6. Insert keyTab($d_1$);
7. End IF;
8. End Loop;
9. For J = 1 to Y Loop// on KeyTab
10. Insert into $ListMapCode_j$
11. Select $MapCode_1, MapCode_2,...., MapCode_m$
12. From MdTab a, KeyTab b
13. Where a.($d_1$) = b.($d_1$)
14. And b.($d_1$) = $kaytab_j.(d_1)$ ;
15. Insert MdTabProces($kaytab_j.(d_1)$ . $ListMapCode_j$ );
16. EndLoop;
17. FI_Gen(MdTabProcess,TmpLargeTab(List of ComDim,Level,Sup),MinSup);
18. Transform_MapCode(TmpLargeTab,MapTab,LargeTab);

In FI_Gen candidate itemsets are generated by equivalence classes[10] and the searching method for candidate itemsets is similar to Apriori algorithm.

After discovering all the large itemsets in the table LargeTab, we will have our hybrid association rule template as follows:

$d_1(val), d_2(val),......, d_m(val) \rightarrow$
$d_2(val)........., d_m(val)$

*4.3 Mining of Association rules*

The mining of association rules is usually a two phase's process. The first phase is for frequent itemsets generation. The second phase generates the rules using another user defined parameter ***minconf,*** which again affects the generation of rules. The second phase is easier and the overall performance of mining association rules is determined mainly by the first step [1].

V. EXPERIMENTAL RESULT

To evaluate the efficiency of the proposed method, the RSHAR, along with the Apriori algorithm, is implemented at the same condition. We use a sample sales database which contains three dimensions (i.e. customer dimension, product, dimension, Promotions dimension) and one sales fact table (see table 3).We perform our experiments using a Pentium IV 1,8 Gigahertz CPU with 512MB.

Table 3. Sales Database

| Table Name | Records |
|---|---|
| Customer Dimension | 100 |
| Product Dimension | 50 |
| Times Dimension | 50 |
| Channel Dimension | 60 |

The minimum support of Apriori algorithm is 0.45%, and the computation times and the numbers of frequent itemsets found by the two algorithms are shown in Figure 2.

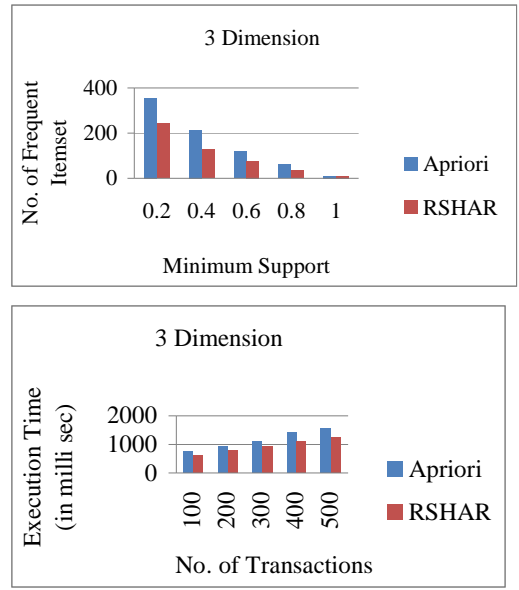

Figure 2. (a) No of frequent itemset. (b)Computation Time

The experimental results in Figure 2 show that the RSHAR performs batter and more rapid than the Apriori algorithm. The RSHAR is not only eliminating considerable amounts of data, but also decreasing the numbers of database scanning, thus reducing the computation quantities to perform data contrasts and also memory requirements.

VI. CONCLUSION

In this paper, the RSHAR is proposed to mining of hybrid association rules. Mining rules with the RSHAR algorithm is two step processes: First we apply the CombineDims *algorithm* to combine the selected dimensions in order to provide the framework for mining hybrid association rules. Then, we apply the GenFI algorithm to discover frequent itemsets in the transaction database. For the new information system, the searching of frequent itemsets is easy based on the concept of equivalence class. The algorithm provides better performance improvements. The gap between the RSHAR and Apriori algorithms becomes evident with the number and size of patterns identified and the searching time reduced. In this paper, we still restricted our proposed extended method to generate association rules on three dimension. In future we will incorporate several dataset and more than three dimensions for mining of hybrid association rules

AUTHORS PROFILE

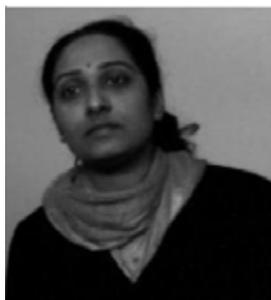

Anjana Pandey, was born on Dec'18, 1978. She completed her Master in Computer Application. Her special fields of interest included Data mining. Presently she is perusing PhD. In MCA Deptt from MANIT, Bhopal.

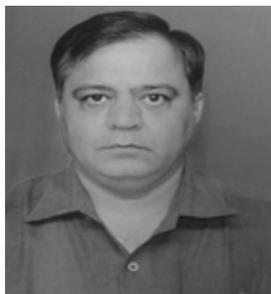

K..R.Pardasani was born on 13th September 1960 at Mathure, India. He completed his graduation, post graduation and PhD (mathematics) from Jiwaji university gwaliar India. his employment experience includes Jiwaji university Gwaliar MDI, Gurgaon and MANIT Bhopal India. Presently he is professor & Head of mathematics at MANIT, Bhopal and his current interest are data mining and computational biology.